\documentclass{article}
\usepackage{amsfonts}

\begin{document}

\title{Pricing of basket options I}
\author{Alexander Kushpel \\
Department of Mathematics, \\
University of Leicester, LE1 7RH, UK\\
E-mail: ak412@le.ac.uk}
\date{10 September 2013}
\maketitle

\begin{abstract}
Pricing of high-dimensional options is a deep problem of the Theoretical
Financial Mathematics. In this article we give a transparent and self
contained treatment of this problem. Namely, we present and study a new
class of L\'{e}vy driven models of stock markets. In our opinion, any market
model should be based on a transparent and intuitively easily acceptable
pre-axiomatic concept. In our case this is the system of stochastic
equations (\ref{stochast}). Our market model is based on the principle of
inheritance, i.e. for the particular choice of parameters it coincides with
known models. Also, our model is effectively numerically realisable. For the
class of models proposed, we give an explicit representations of
characteristic functions. This allows us us to construct a sequence of
approximation formulas to price basket options. We show that our
approximation formulas have almost optimal rate of convergence in the sense
of respective $n$-widths.
\end{abstract}

Keywords: approximation, L\'{e}vy driven models, Fourier transform, reconstruction.

Subject: 91G20, 60G51, 91G60, 91G80.

\section{Introduction}

\bigskip Consider a frictionless market with no arbitrage opportunities and
a constant riskless interest rate $r>0$. Let $S_{j,t}$, $1\leq j\leq n,t\geq
0$, be $n$ asset price processes. Consider a European call option on the
price spread $S_{1,T}-\sum_{j=1}^{n}S_{j,T}$. The common spread option with
maturity $T>0$ and strike $K\geq 0$ is the contract that pays $\left(
S_{1,T}-\sum_{j=1}^{n}S_{j,T}-K\right) _{+}$ at time $T$, where $\left(
a\right) _{+}:=\max \left\{ a,0\right\} $. There is a wide range of such
options traded across different sectors of financial markets. For instance,
the crack spread and crush spread options in the commodity markets \cite%
{Mbanefo}, \cite{Shimko}, credit spread options in the fixed income markets,
index spread options in the equity markets \cite{Duan} and the spark
(fuel/electricity) spread options in the energy markets \cite{Deng}, \cite%
{Pilipovic}.

Assuming the existence of a risk-neutral equivalent martingale measure we
get the following pricing formula for the value at time $0$,

\[
V=e^{-rT}\mathbb{E}\left[ \varphi \right] ,
\]%
where $\varphi $ is a reward function and the expectation is taken with
respect to the equivalent martingale measure. Usually, the reward function
has a simple structure. In particular, in the case of call option,
\[
\varphi =\left( S_{1,T}-\sum_{j=1}^{n}S_{j,T}-K\right) _{+}
\]%
Hence the main problem is to approximate properly the respective density
function and then to approximate $\mathbb{E}\left[ \varphi \right] $. There
is an extensive literature on spread options and their applications. In
particular, if $K=0$ a spread option is the same as an option to exchange
one asset for another. An explicit solution in this case has been obtained
by Margrabe \cite{Margrabe}. Margrabe's model assumes that $S_{1,t}$ and $%
S_{2,t}$ follow a geometric Brownian motion whose volatilities $\sigma _{1}$
and $\sigma _{2}$ do not need to be constant, but the volatility $\sigma $
of $S_{1,t}/S_{2,t}$ is a constant, $\sigma =\left( \sigma _{1}^{2}+\sigma
_{2}^{2}-2\sigma _{1}\sigma _{2}\rho \right) ,$ where $\rho $ is the
correlation coefficient of the Brownian motions $S_{1,t}$ and $S_{2,t}$.
Margrabe's formula states that

\[
V=e^{-q_{1}T}S_{1,0}N\left( d_{1}\right) -e^{-q_{2}T}S_{2,0}N\left(
d_{2}\right) ,
\]%
where $N$ denotes the cumulative distribution for a standard Normal
distribution,
\[
d_{1}=\frac{1}{\sigma T^{1/2}}\left( \ln \left( \frac{S_{1,0}}{S_{2,0}}%
\right) +\left( q_{1}-q_{2}+\frac{\sigma }{2}\right) T\right)
\]%
and $d_{2}=d_{1}-\sigma T^{1/2}$.

Unfortunately, in the case where $K>0$ and $S_{1,t}$, $S_{2,t}$ are
geometric Brownian motions, no explicit pricing formula is known. In this
case various approximation methods have been developed. There are three main
approaches: Monte Carlo techniques which are most convenient for
high-dimensional situation because the convergence is independent of the
dimension, fast Fourier transform methods studied in \cite{Carr and Madan}
and PDEs. Observe that PDE based methods are suitable if the dimension of
the PDE is low (see, e.g. \cite{Jitse Niesen}, \cite{Duffy}, \cite{Tavella}
and \cite{Wilmott} for more information). The usual PDE's approach is based
on numerical approximation resulting in a large system of ordinary
differential equations which can then be solved numerically.

Approximation formulas usually allow quick calculations. In particular, a
popular among practitioners Kirk formula \cite{Kirk} gives a good
approximation to the spread call (see also Carmona-Durrleman procedure \cite%
{carmona}, \cite{Li Deng Zhou}). Various applications of fast Fourier
transform have been considered in \cite{Dempster and Hong} and \cite{Lord}.

It is well-known that Merton-Black-Scholes theory becomes much more
efficient if additional stochastic factors are introduced. Consequently, it
is important to consider a wider family of L\'{e}vy processes. Stable L\'{e}%
vy processes have been used first in this context by Mandelbrot \cite{man1}
and Fama \cite{f1}.

From the 90th L\'{e}vy processes became very popular (see, e.g., \cite{ms1},
\cite{ms2}, \cite{bp1}, \cite{bl1} and references therein).

\bigskip

\section{High-dimensional L\'{e}vy driven models}

In this section we introduce a class of stochastic systems to model
multidimensional return processes.

Let $X_{1,t},\cdot \cdot \cdot ,X_{n,t}$ and $Z_{1,t},\cdot \cdot \cdot
,Z_{n,t}$ be independent random variables, with the densities functions $%
f_{1,t}\left( x_{1}\right) ,\cdot \cdot \cdot ,f_{n,t}\left( x_{n}\right) $
and $z_{1,t}\left( x_{1}\right) ,\cdot \cdot \cdot ,z_{n,t}\left(
x_{n}\right) $ and characteristic exponents $\psi _{s}$ and $\phi _{m},1\leq
s,m\leq n$ respectively. Let $\mathbf{X}_{t}=\left( X_{1,t},\cdot \cdot
\cdot ,X_{n,t}\right) ^{T}$, $\mathbf{Z}_{t}=\left( Z_{1,t},\cdot \cdot
\cdot ,Z_{n,t}\right) ^{T}$ and $\mathrm{A}=\left( a_{j,k}\right) $ be a
real matrix of size $n\times n$. Consider random vector $\mathbf{U}%
_{t}=\left( U_{1,t},\cdot \cdot \cdot ,U_{n,t}\right) ^{T},$
\begin{equation}
\mathbf{U}_{t}=\mathbf{X}_{t}+\mathrm{A}\mathbf{Z}_{t}.  \label{system}
\end{equation}

A matrix $\mathrm{A}$ reflects dependence between the return processes $%
U_{1,t},\cdot \cdot \cdot ,U_{n,t}$ in our model. Assume for simplicity that
$\mathbb{E}\left[ X_{s,t}\right] =0$ and $\mathbb{E}\left[ Z_{s,t}\right]
=0,1\leq s\leq n$. It is easy to check that for any $s$ and $l$, $1\leq
s\neq l\leq n$ the correlation coefficient $\rho \left(
U_{s,t},U_{l,t}\right) $ between $U_{s,t}$ and $U_{l,t}$, where
\[
U_{s,t}=X_{s,t}+\sum_{k=1}^{n}a_{s,k}Z_{s,t},U_{l,t}=X_{l,t}+%
\sum_{k=1}^{n}a_{l,k}Z_{l,t}
\]%
is%
\[
\rho \left( U_{s,t},U_{l,t}\right) =\frac{\mathbb{E}\left[ U_{s,t}U_{l,t}%
\right] }{\mathbb{E}\left[ U_{s,t}^{2}\right] \mathbb{E}\left[ U_{l,t}^{2}%
\right] }
\]%
\[
=\frac{\sum_{k=1}^{n}a_{s,k}^{2}\mathrm{var}\left( Z_{s,t}\right) }{\left(
\left( \mathrm{var}\left( X_{s,t}\right) +\sum_{k=1}^{n}a_{s,k}^{2}\mathrm{%
var}\left( Z_{s,t}\right) \right) \left( \mathrm{var}\left( X_{l,t}\right)
+\sum_{k=1}^{n}a_{l,k}^{2}\mathrm{var}\left( Z_{l,t}\right) \right) \right)
^{1/2}}.
\]%
In particular, if $\mathrm{var}\left( X_{s,t}\right) =\mathrm{var}\left(
Z_{s,t}\right) =v$ and $a_{s,k}=1,1\leq s,k\leq n$ then $\rho \left(
U_{s,t},U_{l,t}\right) =n(n+1)^{-1}$. It reflects our empirical experience:
if the market is in crisis then the prices of stocks are highly correlated.

The next statement gives us an explicit form of the characteristic function
of the return process $\mathbf{U}_{t}.$

{\bf Theorem 1}
\begin{em}
Let $\mathbf{U}_{t}=\mathbf{X}_{t}+\mathrm{A}\mathbf{Z}_{t},$ $\mathrm{A}%
=\left( a_{m,k}\right) $ then in our notations the characteristic function $%
\Phi \left( \mathbf{v,}t\right) $ of $\ \mathbf{U}_{t}$ has the form%
\[
\Phi \left( \mathbf{v,}t\right) =\left( 2\pi \right) ^{n}\left(
\prod\limits_{s=1}^{n}\mathbf{F}^{-1}\left( f_{s,t}\right) \right) \left(
v_{s}\right) \cdot \mathbf{F}^{-1}\left(
\prod\limits_{m=1}^{n}z_{m,t}\right) \left( \mathrm{A}^{\ast }\mathbf{v}%
\right) ,
\]
\[
=\prod\limits_{s=1}^{n}\exp \left( -t\psi _{s}\left( v_{s}\right) \right)
\cdot \prod\limits_{m=1}^{n}\exp \left( -t\phi _{m}\left(
\sum_{k=1}^{n}a_{k,m}v_{k}\right) \right) ,
\]%
where $\mathrm{A}^{\ast }=\left( a_{k,m}\right) $ is the conjugate of $%
\mathrm{A}$.
\end{em}

{\bf Proof} Consider transformation $\mathbb{R}^{2n}\rightarrow \mathbb{R}^{2n}$
defined as
\begin{equation}
\begin{array}{c}
\mathbf{U}_{t}=\mathbf{X}_{t}+\mathrm{A}\mathbf{Z}_{t}, \\
\mathbf{Z}_{t}=\mathbf{Z}_{t}.%
\end{array}
\label{stochast}
\end{equation}%
Hence the inverse is given by
\[
\begin{array}{c}
\mathbf{X}_{t}=\mathbf{U}_{t}-\mathrm{A}\mathbf{Z}_{t}, \\
\mathbf{Z}_{t}=\mathbf{Z}_{t}.%
\end{array}%
\]%
or
\[
\left(
\begin{array}{c}
\mathbf{X}_{t} \\
\mathbf{Z}_{t}%
\end{array}%
\right) =\left(
\begin{array}{cc}
\mathrm{I} & -\mathrm{A} \\
\mathbf{0} & \mathrm{I}%
\end{array}%
\right) \left(
\begin{array}{c}
\mathbf{U}_{t} \\
\mathbf{Z}_{t}%
\end{array}%
\right)
\]%
and the Jacobian $J$ of this transform is
\[
J=\det \left(
\begin{array}{cc}
\mathrm{I} & -\mathrm{A} \\
\mathbf{0} & \mathrm{I}%
\end{array}%
\right) =1,
\]%
where $\mathrm{I}=\mathrm{I}_{n\times n}$ is an identity. The density
function $\phi _{t}\left( \mathbf{u}_{t},\mathbf{z}_{t}\right) $ is given by
\[
\phi _{t}\left( \mathbf{u},\mathbf{z}\right)
=\prod\limits_{s=1}^{n}f_{s,t}\left(
u_{s}-\sum_{m=1}^{n}a_{s,m}z_{m}\right) \prod\limits_{l=1}^{n}z_{l,t}\left(
z_{l}\right) .
\]%
It means that the density function $\omega _{t}\left( \mathbf{u}\right) $ is
\[
\omega _{t}\left( \mathbf{u}\right) =\int_{\mathbb{R}^{n}}\phi _{t}\left(
\mathbf{u},\mathbf{z}\right) d\mathbf{z}
\]%
and the characteristic function has the form%
\[
\Phi \left( \mathbf{v,}t\right) :=\mathbb{E}\left[ \exp \left( i\left\langle
\mathbf{U}_{t},\mathbf{v}\right\rangle \right) \right] :=\exp \left( -t\psi
\left( \mathbf{v}\right) \right) =\mathbf{F}\omega _{t}\left( \mathbf{v}%
\right)
\]%
\[
=\int_{\mathbb{R}^{n}}\exp \left( i\left\langle \mathbf{u},\mathbf{v}%
\right\rangle \right) \omega _{t}\left( \mathbf{u}\right) d\mathbf{u}
\]%
\[
=\int_{\mathbb{R}^{n}}\exp \left( i\left\langle \mathbf{u},\mathbf{v}%
\right\rangle \right) \left( \int_{\mathbb{R}^{n}}\phi _{t}\left( \mathbf{u},%
\mathbf{z}\right) d\mathbf{z}\right) d\mathbf{u}
\]%
\[
=\int_{\mathbb{R}^{n}}\exp \left( i\left\langle \mathbf{u},\mathbf{v}%
\right\rangle \right) \left( \int_{\mathbb{R}^{n}}\prod%
\limits_{s=1}^{n}f_{s,t}\left( u_{s}-\sum_{m=1}^{n}a_{s,m}z_{m}\right)
\prod\limits_{m=1}^{n}z_{m,t}\left( z_{m}\right) d\mathbf{z}\right) d%
\mathbf{u}
\]%
\begin{equation}
=\int_{\mathbb{R}^{n}}\left( \prod\limits_{s=1}^{n}\int_{\mathbb{R}%
}f_{s,t}\left( u_{s}-\sum_{m=1}^{n}a_{s,m}z_{m}\right) \exp \left(
iu_{s}v_{s}\right) du_{s}\right) \prod\limits_{m=1}^{n}z_{m,t}\left(
z_{m}\right) d\mathbf{z.}  \label{2013}
\end{equation}%
Let $\xi _{s}=u_{s}-\sum_{m=1}^{n}a_{s,m}z_{m},1\leq s\leq n$ then
\[
\int_{\mathbb{R}}f_{s,t}\left( u_{s}-\sum_{m=1}^{n}a_{s,m}z_{m}\right) \exp
\left( iu_{s}v_{s}\right) du_{s}
\]%
\[
=\int_{\mathbb{R}}f_{s,t}\left( \xi _{s}\right) \exp \left( i\left( \xi
_{s}+\sum_{m=1}^{n}a_{s,m}z_{m}\right) v_{s}\right) d\xi _{s}
\]%
\[
=\exp \left( iv_{s}\sum_{m=1}^{n}a_{s,m}z_{m}\right) \int_{\mathbb{R}%
}f_{s,t}\left( \xi _{s}\right) \exp \left( i\xi _{s}v_{s}\right) d\xi _{s}
\]%
\begin{equation}
=\exp \left( iv_{s}\sum_{m=1}^{n}a_{s,m}z_{m}\right) 2\pi \mathbf{F}%
^{-1}\left( f_{s,t}\right) \left( v_{s}\right)  \label{2014}
\end{equation}%
Comparing \ref{2013} and \ref{2014} we get
\[
\Phi \left( \mathbf{v,}t\right) =\int_{\mathbb{R}^{n}}\left(
\prod\limits_{s=1}^{n}\exp \left( iv_{s}\sum_{m=1}^{n}a_{s,m}z_{m}\right)
2\pi \mathbf{F}^{-1}\left( f_{s,t}\right) \left( v_{s}\right) \right)
\prod\limits_{m=1}^{n}z_{m,t}\left( z_{m}\right) d\mathbf{z}
\]%
\[
=\prod\limits_{s=1}^{n}2\pi \mathbf{F}^{-1}\left( f_{s,t}\right) \left(
v_{s}\right) \int_{\mathbb{R}^{n}}\left( \prod\limits_{s=1}^{n}\exp \left(
iv_{s}\sum_{m=1}^{n}a_{s,m}z_{m}\right) \right)
\prod\limits_{m=1}^{n}z_{m,t}\left( z_{m}\right) d\mathbf{z}
\]%
\[
=\prod\limits_{s=1}^{n}2\pi \mathbf{F}^{-1}\left( f_{s,t}\right) \left(
v_{s}\right) \int_{\mathbb{R}^{n}}\exp \left( i\sum_{s=1}^{n}\left(
v_{s}\sum_{m=1}^{n}a_{s,m}z_{m}\right) \right)
\prod\limits_{m=1}^{n}z_{m,t}\left( z_{m}\right) d\mathbf{z}
\]%
\[
=\prod\limits_{s=1}^{n}2\pi \mathbf{F}^{-1}\left( f_{s,t}\right) \left(
v_{s}\right) \int_{\mathbb{R}^{n}}\exp \left( \left\langle \mathbf{v},%
\mathrm{A}\mathbf{z}\right\rangle \right) \left(
\prod\limits_{m=1}^{n}z_{m,t}\left( z_{m}\right) \right) d\mathbf{z}
\]%
\[
=\prod\limits_{s=1}^{n}2\pi \mathbf{F}^{-1}\left( f_{s,t}\right) \left(
v_{s}\right) \int_{\mathbb{R}^{n}}\exp \left( \left\langle \mathrm{A}^{\ast }%
\mathbf{v},\mathbf{z}\right\rangle \right) \left(
\prod\limits_{m=1}^{n}z_{m,t}\left( z_{m}\right) \right) d\mathbf{z}
\]%
\[
=\prod\limits_{s=1}^{n}2\pi \mathbf{F}^{-1}\left( f_{s,t}\right) \left(
v_{s}\right) \cdot \mathbf{F}^{-1}\left( \prod\limits_{m=1}^{n}2\pi
z_{m,t}\right) \left( \mathrm{A}^{\ast }\mathbf{v}\right)
\]%
\[
=\prod\limits_{s=1}^{n}2\pi \mathbf{F}^{-1}\left( f_{s,t}\right) \left(
v_{s}\right) \cdot \mathbf{F}^{-1}\left( \prod\limits_{m=1}^{n}2\pi
z_{m,t}\right) \left( \mathrm{A}^{\ast }\mathbf{v}\right)
\]%
\[
=\prod\limits_{s=1}^{n}2\pi \mathbf{F}^{-1}\left( f_{s,t}\right) \left(
v_{s}\right) \cdot \mathbf{F}^{-1}\left( \prod\limits_{m=1}^{n}2\pi
z_{m,t}\right) \left( \mathrm{A}^{\ast }\mathbf{v}\right) ,
\]
where $\mathrm{A}^{\ast }=\left( a_{k,j}\right) $ is a conjugate to $\mathrm{%
A}$. Hence
\[
\Phi \left( \mathbf{v,}t\right) =\prod\limits_{s=1}^{n}\exp \left( -t\psi
_{s}\left( v_{s}\right) \right) \cdot \prod\limits_{m=1}^{n}\exp \left(
-t\phi _{m}\left( \sum_{k=1}^{n}a_{k,m}v_{k}\right) \right) .
\]

\section{The equivalent martingale measure condition}

In this section we specify an equivalent martingale measure condition for
our model. Under the equivalent martingale measure all assets have the same
expected rate of return which is a risk free rate. It simply means that
under no-arbitrage conditions the risk preferences of investors acting on
the market do not enter into valuation decisions. Consider a frictionless
market consisting of a riskless bond $B$ and stock $S$. In this market $S$
is modeled by an exponential L\'{e}vy process $S=S_{t}=S_{0}e^{X_{t}}$ under
a chosen equivalent martingale measure $\mathbb{Q}$. Assume that the
riskless rate $r$ is constant. The next statement is a generalisation of a
known result. In the previous versions authors assumed that the
characteristic exponent $\psi $ admits an analytic extension into the strip $%
\left\{ z\left\vert -1\leq {\rm Im}z\leq 0\right. \right\} $ (see e.g. \cite%
{bl1}).

{\bf Theorem 2. }
\begin{em}
Let $\mathbb{Q}$ be a chosen equivalent
martingale measure and $D\subset \mathbb{R+}i\mathbb{R}$ be the domain of
the characteristic exponent $\psi ^{\mathbb{Q}}$. Assume that $\mathbb{R\cup
}\left\{ -i\right\} \subset D$, then in our notations $\psi ^{\mathbb{Q}%
}\left( -i\right) =-r$.
\end{em}

{\bf Proof}
The discount price process which is given by
\[
\widetilde{S}_{t}=\exp \left( -rt\right) S_{t}=\exp \left( -rt\right)
S_{0}\exp \left( X_{t}\right)
\]%
must be a martingale under a chosen equivalent martingale measure $\mathbb{Q}
$, i.e. for any $0\leq l<t\leq T$ the martingale condition must hold,
\[
\widetilde{S}_{l}=\mathbb{E}^{\mathbb{Q}}\left[ \widetilde{S}_{t}\left\vert
\mathcal{F}_{l}\right. \right] .
\]%
In particular, let $l=0$ then for any $t\in \left( 0,T\right] $ we have
\[
\widetilde{S}_{0}=S_{0}\exp \left( -r0\right) =S_{0}=\mathbb{E}^{\mathbb{Q}}%
\left[ S_{0}\exp \left( -rt\right) \exp \left( X_{t}\right) \left\vert
\mathcal{F}_{0}\right. \right]
\]%
\[
=\mathbb{E}^{\mathbb{Q}}\left[ S_{0}\exp \left( -rt\right) \exp \left(
X_{t}\right) \right] =S_{0}\mathbb{E}^{\mathbb{Q}}\left[ \exp \left(
-rt\right) \exp \left( X_{t}\right) \right] .
\]%
Since $S_{0}>0$ then $\mathbb{E}^{\mathbb{Q}}\left[ \exp \left( -rt\right)
\exp \left( X_{t}\right) \right] =1$. Let $t=T$ then $\exp \left( rT\right) =%
\mathbb{E}^{\mathbb{Q}}\left[ \exp \left( X_{T}\right) \right] $. Since $%
-i\in D$ then by the definition of the characteristic exponent
\begin{equation}
\exp \left( -T\psi ^{\mathbb{Q}}\left( -i\right) \right) =\mathbb{E}^{%
\mathbb{Q}}\left[ \exp \left( i\left( -i\right) X_{t}\right) \right] =%
\mathbb{E}^{\mathbb{Q}}\left[ \exp \left( X_{t}\right) \right] .
\label{exp1}
\end{equation}%
Hence, since $T>0$ then from (\ref{exp1}) it follows that $r=-\psi ^{\mathbb{%
Q}}\left( -i\right) .$

In general $\mathbb{Q}$ is not unique. In what follows we assume that $%
\mathbb{Q}$ has been fixed and all expectations will be computed with
respect to this measure.

We specify now the equivalent martingale measure condition for the system (%
\ref{system}).

{Theorem 3. }
\begin{em}
Let the stock prices are modeled by%
\[
S_{s,t}=S_{s,0}\exp \left( U_{s,t}\right) ,1\leq s\leq n.
\]%
and the domain $D\subset \mathbb{R}^{n}\mathbb{+}i\mathbb{R}^{n}$ of the
characteristic exponent $\psi ^{\mathbb{Q}}$ contains $\mathbb{R}^{n}\cup
\left( \cup _{k=1}^{n}\left\{ -i\mathbf{e}_{k}\right\} \right) $\ where\ $%
\left\{ \mathbf{e}_{k},1\leq k\leq n\right\} $\ is the standard basis in \ $%
\mathbb{R}^{n}$ then \
\[
\psi ^{\mathbb{Q}}\left( -i\mathbf{e}_{s}\right) =-r,1\leq s\leq n.
\]
\end{em}

{\bf Proof}
Observe that for any $1\leq s\leq n$ the discount price process $S_{s,t}$
must be a martingale under a chosen equivalent martingale measure $\mathbb{Q}%
.$ Let $\psi _{s}^{\mathbb{Q}}\left( x_{s}\right) $ be the characteristic
exponent of $U_{s,t}$ then
\[
\exp \left( -t\psi _{s}^{\mathbb{Q}}\left( x_{s}\right) \right) =\mathbb{E}^{%
\mathbb{Q}}\left[ \exp \left( \left\langle ix_{s}\mathbf{,}%
U_{s,t}\right\rangle \right) \right]
\]%
\[
=\mathbb{E}^{\mathbb{Q}}\left[ \exp \left( \left\langle i\mathbf{x,}U_{s,t}%
\mathbf{e}_{s}\right\rangle \right) \right] =\exp \left( -t\psi ^{\mathbb{Q}%
}\left( x_{s}\mathbf{e}_{s}\right) \right)
\]%
and by the Theorem 2 we get $r=-\psi _{s}^{\mathbb{Q}}\left( -i\right) $
which gives a system of $n$ equations
\[
\psi ^{\mathbb{Q}}\left( -i\mathbf{e}_{s}\right) =-r,1\leq s\leq n.
\]

Observe that in general riskless interest rate may depend on $s$. In this
case we get the system $\psi ^{\mathbb{Q}}\left( -i\mathbf{e}_{s}\right)
=-r_{s},1\leq s\leq n.$

\section{KoBoL family}

In this section we study characteristic exponents of KoBoL family. The idea
is based on a simple observation. From the L\'{e}vi-Khintchine formula (\ref%
{Levy-Khintchine1}) it follows that it is possible to find $\psi \left( \xi \right)
$ explicitly if we can find the inverse Fourier transform \ of $\Pi \left(
dx\right) .$ It was suggested by the authors of \cite{bl1} to consider the
following form of \ $\Pi \left( dx\right) ,$%
\[
\Pi \left( dx\right) =\left\vert x\right\vert ^{\alpha }\exp \left( -\beta
\left\vert x\right\vert \right) ,
\]%
where $\alpha $ and $\beta $ are fixed parameters.

A known class of high-dimensional models is based on so-called KoBoL family
which is given by%
\[
\Pi \left( d\mathbf{x}\right) =\rho ^{-\nu -1}\exp \left( -\lambda \left(
\mathbf{\phi }\right) \rho \right) d\rho \Pi ^{^{\prime }}\left( d\mathbf{%
\phi }\right) ,
\]%
where $\Pi ^{^{\prime }}\left( d\mathbf{\phi }\right) $ is a finite measure
on the unit sphere $\mathbb{S}^{n-1}$ and $\lambda :C\left( \mathbb{S}%
^{n-1}\right) \rightarrow \mathbb{R}_{+}$ \cite{bl1}. The respective
characteristic exponent has the form
\[
\psi \left( \mathbf{\xi }\right) =-i\left\langle \mathbf{\mu ,\xi }%
\right\rangle +\Gamma \left( -\nu \right) \int_{\mathbb{S}^{n-1}}\left(
\left( \lambda \left( \mathbf{\phi }\right) \right) ^{\nu }-\left( \lambda
\left( \mathbf{\phi }\right) -i\left\langle \Sigma \mathbf{\xi ,\phi }%
\right\rangle \right) ^{\nu }\right) \Pi ^{^{\prime }}\left( d\mathbf{\phi }%
\right) ,
\]%
where $\nu \in \left( 0,2\right) ,\mathbf{\mu \in }\mathbb{R}^{n}$ and $%
\Sigma $ is a positive-definite matrix. Clearly%
\[
\psi \left( \mathbf{\xi }\right) =-i\left\langle \mathbf{\mu ,\xi }%
\right\rangle +C_{1}-C_{2}\left( \mathbf{\xi }\right) ,
\]%
where
\[
C_{1}:=\int_{\mathbb{S}^{n-1}}\left( \lambda \left( \mathbf{\phi }\right)
\right) ^{\nu }\Pi ^{^{\prime }}\left( d\mathbf{\phi }\right)
\]%
and
\[
C_{2}\left( \mathbf{\xi }\right) :=\int_{\mathbb{S}^{n-1}}\left( \lambda
\left( \mathbf{\phi }\right) -i\left\langle \Sigma \mathbf{\xi ,\phi }%
\right\rangle \right) ^{\nu }\Pi ^{^{\prime }}\left( d\mathbf{\phi }\right)
.
\]%
Let in particular $\Pi ^{^{\prime }}\left( d\mathbf{\phi }\right) =cd\mathbf{%
\phi ,}$ where $c>0$ and $d\mathbf{\phi }$ is the Haar measure on $\mathbb{S}%
^{n-1}$. Then the problem is to approximate the integral
\[
C_{2}\left( \mathbf{\xi }\right) :=c\int_{\mathbb{S}^{n-1}}\left( \lambda
\left( \mathbf{\phi }\right) -i\left\langle \Sigma \mathbf{\xi ,\phi }%
\right\rangle \right) ^{\nu }d\mathbf{\phi }.
\]%
This problem is computationally difficult. In this section we construct a
class of KoBoL processes which are based on the respective one-dimensional
blocks. This allows us to simplify the expression of the characteristic
exponent.

We start with a one-dimensional version of the Theorem 5,%
\[
\psi \left( \xi \right) =2^{-1}a\xi ^{2}-i\gamma \xi -\int_{\mathbb{R}%
}\left( \exp \left( ix\xi \right) -1-ix\xi \chi _{\left[ -1,1\right] }\left(
x\right) \right) \Pi \left( dx\right) ,
\]%
where $a\geq 0,\gamma \in \mathbb{R}$ and $\Pi $ is a measure on $\mathbb{R}$
satisfying%
\[
\Pi \left( \left\{ 0\right\} \right) =0,\int_{\mathbb{R}}\min \left\{
x^{2},1\right\} \Pi \left( dx\right) <\infty .
\]%
Let $a=\gamma =0,0<\nu <2,\lambda >0,$%
\[
\Pi ^{+}\left( \nu ,\lambda ,dx\right) =x_{+}^{-\nu -1}\exp \left( -\lambda
x\right) dx,
\]%
\[
\Pi ^{-}\left( \nu ,\lambda ,dx\right) =x_{-}^{-\nu -1}\exp \left( \lambda
x\right) dx,
\]%
where $x_{+}=\max \left\{ x,0\right\} ,x_{-}=x_{+}-x$ and
\begin{equation}
\Pi \left( dx\right) =c_{+}\Pi ^{+}\left( \nu ,-\lambda _{-},dx\right)
+c_{-}\Pi ^{-}\left( \nu ,\lambda _{+},dx\right) ,c_{+}>0,c_{-}>0,\lambda
_{-}<0<\lambda _{+}.  \label{measureL2}
\end{equation}%
It is easy to check that
\[
\int_{\mathbb{R}}\min \left\{ x^{2},1\right\} \left( c_{+}\Pi ^{+}\left( \nu
,-\lambda _{-},dx\right) +c_{-}\Pi ^{-}\left( \nu ,\lambda _{+},dx\right)
\right) <\infty .
\]%
Hence (\ref{measureL2}) defines a L\'{e}vy measure. Moreover, if $\nu <1$
then
\[
\int_{\mathbb{R}}\min \left\{ \left\vert x\right\vert ,1\right\} \left(
c_{+}\Pi ^{+}\left( \nu ,-\lambda _{-},dx\right) +c_{-}\Pi ^{-}\left( \nu
,\lambda _{+},dx\right) \right) <\infty
\]%
and the process has a finite variation.

Lemmas 3.1 and 3.2 \cite{bl1} give a representation of the respective
characteristic exponent%
\[
\psi \left( \xi \right) =-i\mu \xi +c_{+}\Gamma \left( -\nu \right) \left(
\left( -\lambda _{-}\right) ^{\nu }-\left( -\lambda _{-}-i\xi \right) ^{\nu
}\right)
\]%
\begin{equation}
+c_{-}\Gamma \left( -\nu \right) \left( \lambda _{+}^{\nu }-\left( \lambda
_{+}+i\xi \right) ^{\nu }\right) ,\nu \in \left( 0,1\right) \cup \left(
1,2\right) .  \label{representation-k}
\end{equation}%
The proof of Lemma 3.2 presented in \cite{bl1} is incomplete. The next
statement gives a complete proof of the representation (\ref%
{representation-k}) which is important in our applications.

{Theorem 4. }
\begin{em}
Let $\nu \in \left( 0,1\right) $ then in our notations
\[
\psi \left( \xi \right) =-i\mu \xi +c_{+}\Gamma \left( -\nu \right) \left(
\left( -\lambda _{-}\right) ^{\nu }-\left( -\lambda _{-}-i\xi \right) ^{\nu
}\right)
\]%
\[
+c_{-}\Gamma \left( -\nu \right) \left( \lambda _{+}^{\nu }-\left( \lambda
_{+}+i\xi \right) ^{\nu }\right) ,
\]%
where $\mu $ is a real parameter.
\end{em}

{\bf Proof}
It is sufficient to prove the statement just for the $\Pi ^{+}\left( \nu
,\lambda ,dx\right) $, i.e. to find
\[
-\psi ^{+}\left( \xi \right) :=\int_{\mathbb{R}}\left( \exp \left( ix\xi
\right) -1-ix\xi \chi _{\left[ -1,1\right] }\left( x\right) \right) \Pi
^{+}\left( dx\right)
\]%
\[
=\int_{\mathbb{R}}\left( \exp \left( ix\xi \right) -1-ix\xi \chi _{\left[
-1,1\right] }\left( x\right) \right) x_{+}^{-\nu -1}\exp \left( -\lambda
x\right) dx
\]%
\[
=\int_{0}^{\infty }\left( \exp \left( ix\xi \right) -1-ix\xi \chi _{\left[
-1,1\right] }\left( x\right) \right) x^{-\nu -1}\exp \left( -\lambda
x\right) dx
\]%
\[
=\int_{0}^{\infty }\left( \exp \left( ix\xi \right) -1\right) x^{-\nu
-1}\exp \left( -\lambda x\right) dx
\]%
\[
-i\xi \int_{0}^{1}x^{-\nu }\exp \left( -\lambda x\right) dx
\]%
\[
=\int_{0}^{\infty }\left( \exp \left( ix\xi \right) -1\right) x^{-\nu
-1}\exp \left( -\lambda x\right) dx-i\xi \mathrm{B}\left( \nu ,\lambda
\right)
\]%
\[
:=I_{1}\left( \xi ,\nu ,\lambda \right) -i\xi \mathrm{B}\left( \nu ,\lambda
\right) ,
\]%
where $\mathrm{B}\left( \nu ,\lambda \right) :=\int_{0}^{1}x^{-\nu }\exp
\left( -\lambda x\right) dx$ and
\[
I_{1}\left( \xi ,\nu ,\lambda \right) =-\frac{1}{\nu }\int_{0}^{\infty
}\left( \exp \left( -\left( \lambda -i\xi \right) x\right) -\exp \left(
-\lambda x\right) \right) dx^{-\nu }
\]%
\[
=-\frac{1}{\nu }\left( \left( \exp \left( -\left( \lambda -i\xi \right)
x\right) -\exp \left( -\lambda x\right) \right) x^{-\nu }\right) \left\vert
_{0}^{\infty }\right.
\]%
\[
-\left( -\frac{1}{\nu }\right) \int_{0}^{\infty }\left( -\left( \lambda
-i\xi \right) \exp \left( -\left( \lambda -i\xi \right) x\right) +\lambda
\exp \left( -\lambda x\right) \right) x^{-\nu }dx
\]%
\[
=-\frac{\lambda -i\xi }{\nu }\int_{0}^{\infty }\exp \left( -\left( \lambda
-i\xi \right) x\right) x^{-\nu }dx-\lambda ^{\nu }\Gamma \left( -\nu \right)
:=I_{2}-\lambda ^{\nu }\Gamma \left( -\nu \right) .
\]%
Making change of variable $z=\left( \lambda -i\xi \right) x$ in $I_{2}$ we
get%
\[
I_{2}=-\frac{\left( \lambda -i\xi \right) ^{\nu }}{\nu }\int_{\gamma }\exp
\left( -z\right) z^{-\nu }dz,
\]%
where $\gamma $ is the ray $\left\{ z\left\vert z=\left( \lambda -i\xi
\right) x,\lambda >0,\xi \in \mathbb{R}\right. \right\} $, $\lambda $ and $%
\xi $ are fixed parameters and $x\geq 0$. Assume that $\xi \geq 0$. The case
$\xi \leq 0$ can be treated similarly. Consider the contour $\eta :=\gamma
_{1}\cup \gamma _{2}\cup \gamma _{3}\cup \gamma _{4}$, where
\[
\gamma _{1}:=\left\{ z=\rho \exp \left( i\theta \right) \left\vert 0\leq
\theta \leq \arg \left( \lambda -i\xi \right) ,\lambda >0,\xi \in \mathbb{R}%
\right. \right\} ,
\]%
\[
\gamma _{2}:=\left\{ z\left\vert \rho \leq z\leq R,z\in \mathbb{R}\right.
\right\}
\]%
\[
\gamma _{3}:=\left\{ z=R\exp \left( i\theta \right) \left\vert 0\leq \theta
\leq \arg \left( \lambda -i\xi \right) ,\lambda >0,\xi \in \mathbb{R}\right.
\right\} ,
\]%
\[
\gamma _{4}:=\left\{ z\left\vert z=\left( \lambda -i\xi \right) x,\rho \leq
\left\vert z\right\vert \leq R\right. \right\} .
\]%
The function $\exp \left( -z\right) z^{-\nu }$ is analytic in the domain
bounded by $\eta $, hence from the Cauchy's theorem it follows that
\[
\int_{\eta }\exp \left( -z\right) z^{-\nu }dz=0
\]%
and since $\xi \geq 0$ then for some $\delta >0$ we get $-\pi /2+\delta \leq
\arg \left( \lambda -i\xi \right) \leq 0.$ Hence
\[
\lim_{R\rightarrow \infty }\left\vert \int_{\gamma _{3}}\exp \left(
-z\right) z^{-\nu }dz\right\vert
\]%
\[
=\lim_{R\rightarrow \infty }\left\vert \int_{0}^{\arg \left( \lambda -i\xi
\right) }\exp \left( -R\mathrm{exp}\left( i\theta \right) \right)
R^{-\nu }\exp \left( -i\nu \theta \right) Ri\exp \left( i\theta \right)
d\theta \right\vert
\]%
\[
\leq \frac{\pi }{2}\lim_{R\rightarrow \infty }\exp \left( -R\cos \delta
\right) \exp \left( R^{1-\nu }\right) =0.
\]%
Observe that%
\[
\lim_{\rho \rightarrow 0}\left\vert \int_{\gamma _{3}}\exp \left( -z\right)
z^{-\nu }dz\right\vert
\]%
\[
\leq \lim_{\rho \rightarrow 0}\left\vert \int_{0}^{2\pi }\exp \left( -\rho
\mathrm{exp}\left( i\theta \right) \right) \rho ^{-\nu }\exp \left(
-i\nu \theta \right) \rho i\exp \left( i\theta \right) d\theta \right\vert
\]%
\[
\leq 2\pi \lim_{\rho \rightarrow 0}\rho ^{-\nu +1}=0.
\]%
Hence%
\[
\int_{\gamma }\exp \left( -z\right) z^{-\nu }dz=\int_{\mathbb{R}_{+}}\exp
\left( -z\right) z^{-\nu }dz=\Gamma \left( -\nu +1\right) =-\nu \Gamma
\left( -\nu \right).
\]%
Consequently
\[
I_{2}=-\frac{\left( \lambda -i\xi \right) ^{\nu }}{\nu }\int_{\gamma }\exp
\left( -y\right) y^{-\nu }dy=\Gamma \left( -\nu \right) \left( \lambda -i\xi
\right) ^{\nu }
\]%
and%
\[
\psi ^{+}\left( \xi \right) =\Gamma \left( -\nu \right) \left( \lambda ^{\nu
}-\left( \left( \lambda -i\xi \right) ^{\nu }\right) \right) +i\xi \mathrm{B}%
\left( \nu ,\lambda \right) .
\]

\section{Appendix I: Stochastic processes and density functions}

Let $\mathcal{B}\left( \mathbb{R}^{n}\right) $ be the collection of all
Borel sets in $\mathbb{R}^{n}$ (which is the $\sigma -$algebra generated by
all open sets in $\mathbb{R}^{n}$). A mapping $\mathbf{X}:\mathbb{R}%
^{n}\rightarrow \mathbb{R}^{n}$ is an $\mathbb{R}^{n}$-valued random
variable if it is $\mathcal{B}\left( \mathbb{R}^{n}\right) $ measurable,
i.e. for any $B\in \mathcal{B}\left( \mathbb{R}^{n}\right) $ we have $%
\left\{ \omega \left\vert \mathbf{X}\left( \omega \right) \in B\right.
\right\} \in \mathcal{B}\left( \mathbb{R}^{n}\right) $. Let $\left( \mathbb{R%
}^{n},\mathcal{B}\left( \mathbb{R}^{n}\right) ,\mathrm{P}\right) $ be a
fixed probability space. A stochastic process $\mathbf{X}=\left\{ \mathbf{X}%
_{t},t\in \mathbb{R}\right\} $ is a one-parametric family of random
variables on a common probability space $\left( \mathbb{R}^{n},\mathcal{B}%
\left( \mathbb{R}^{n}\right) ,\mathrm{P}\right) $. The\textit{\ }trajectory
of the process $\mathbf{X}$ is a map
\[
\begin{array}{ccc}
\mathbb{R}_{+} & \longrightarrow & \mathbb{R}^{n} \\
t & \longmapsto & \mathbf{X}_{t}\left( \omega \right) ,%
\end{array}%
\]%
where $\omega \in \Omega $ and $\mathbf{X}_{t}=\left( X_{1t},\cdot \cdot
\cdot ,X_{nt}\right) $.

$\mathbf{X}=\{\mathbf{X}_{t}\}_{t\in \mathbb{R}_{+}}$ is called a L\'{e}vy
process (process with stationary independent increments) if

\begin{enumerate}
\item The random variables $\mathbf{X}_{t_{0}},\mathbf{X}_{t_{1}}-\mathbf{X}%
_{t_{0}},\cdots ,\mathbf{X}_{t_{m}}-\mathbf{X}_{t_{m-1}}$, for any $0\leq
t_{0}<t_{1}<\cdots <t_{m}$ and $m\in \mathbb{N}$ are independent
(independent increment property).

\item $\mathbf{X}_{0}=\mathbf{0}$ a.s.

\item The distribution of $\mathbf{X}_{t+\tau }-\mathbf{X}_{t}$ is
independent of $\tau $ (temporal homogeneity or stationary increments
property).

\item It is stochastically continuous, i.e.
\[
\lim_{\tau \rightarrow t}\mathrm{P}\left[ |\mathbf{X}_{\tau }-\mathbf{X}%
_{t}|>\epsilon \right] =0
\]%
for any $\epsilon >0$ and $t\geq 0$.

\item There is $\Omega _{0}\in \mathcal{F}$ with $\mathrm{P}\left( \Omega
_{0}\right) =1$ such that, for any $\omega \in \Omega _{0},$ $\mathbf{X}%
_{t}\left( \omega \right) $ is right-continuous on $\mathbb{[}0,\infty )$
and has left limits on $\mathbb{(}0,\infty ).$
\end{enumerate}

A process satisfying ($1-4$) is called a L\'{e}vy process in law.
An additive process is a stochastic process which satisfies ($1,2,4,5$) and
an\ additive process in law satisfies ($1,2,4$).

For an integrable on $\mathbb{R}^{n}$ function, $f\in L_{1}\left( \mathbb{R}%
^{n}\right) $ define its Fourier transform
\[
\mathbf{F}f\left( \mathbf{y}\right) =\int_{\mathbb{R}^{n}}\exp \left(
-i\left\langle \mathbf{x,y}\right\rangle \right) f\left( \mathbf{x}\right) d%
\mathbf{x}
\]%
and its formal inverse
\[
\left( \mathbf{F}^{-1}f\right) \left( \mathbf{x}\right) =\left( 2\pi \right)
^{-n}\int_{\mathbb{R}^{n}}\exp \left( i\left\langle \mathbf{x,y}%
\right\rangle \right) f\left( \mathbf{y}\right) d\mathbf{y.}
\]%
Let $\left\langle \mathbf{u,v}\right\rangle :=\sum_{k=1}^{n}u_{k}v_{k}$ be
the canonic scalar product on $\mathbb{R}^{n}$, $\mathbf{u=}\left(
u_{1},\cdot \cdot \cdot ,u_{n}\right) ,\mathbf{v=}\left( v_{1},\cdot \cdot
\cdot ,v_{n}\right) $. The characteristic function of the distribution of $%
\mathbf{X}_{t}$ of any L\'{e}vy process can be represented in the form
\[
\mathbb{E}\left[ \exp \left( \left\langle i\mathbf{x,X}_{t}\right\rangle
\right) \right] =e^{-t\psi \left( \mathbf{x}\right) }
\]%
\[
=\left( 2\pi \right) ^{n}\mathbf{F}^{-1}p_{t}\left( \mathbf{x}\right) ,
\]%
where $p_{t}\left( \mathbf{x}\right) $ is the density function of $\mathbf{X}%
_{t}$, $\mathbf{x}\in \mathbb{R}^{n}$, $t\in \mathbb{R}_{+}$ and the
function $\psi \left( \mathbf{x}\right) $ \ is uniquely determined. This
function is called the characteristic exponent. Vice versa, a L\'{e}vy
process $\mathbf{X}=\{\mathbf{X}_{t}\}_{t\in \mathbb{R}_{+}}$ is determined
uniquely by its characteristic exponent $\psi \left( \mathbf{x}\right) $. In
particular, density function $p_{t}$ can be expressed as
\[
p_{t}\left( \cdot \right) =\left( 2\pi \right) ^{-n}\int_{\mathbb{R}%
^{n}}\exp \left( -i\left\langle \cdot \mathbf{,x}\right\rangle -t\psi \left(
\mathbf{x}\right) \right) d\mathbf{x}=\left( 2\pi \right) ^{-n}\mathbf{F}%
\left( \exp \left( -t\psi \left( \mathbf{x}\right) \right) \right) \left(
\cdot \right) .
\]%
The key role in our analysis plays the following classical result known as
the L\'{e}vy-Khintchine formula which gives a representation of
characteristic functions of all infinitely divisible distributions.

{Theorem 5. }
\begin{em}
 Let $\mathbf{X}=\{\mathbf{X}_{t}\}_{t\in \mathbb{R}_{+}}$
be a L\'{e}vy process on $\mathbb{R}^{n}$. Then its characteristic exponent
admits the representation
\begin{equation} \label{Levy-Khintchine1}
\psi (\mathbf{y})=2^{-1}\left\langle A\mathbf{y},\mathbf{y}\right\rangle
-i\langle \mathbf{b},\mathbf{y}\rangle -\int_{\mathbb{R}^{n}}\left(
e^{i\langle \mathbf{y},\mathbf{x}\rangle }-1-i\langle \mathbf{y},\mathbf{x}%
\rangle \chi _{D}(\mathbf{x})\right) \Pi (d\mathbf{x}),
\end{equation}
where $\chi _{D}(\mathbf{x})$ is the characteristic function of $D:=\{%
\mathbf{x}\in \mathbb{R}^{n},\,\,|\mathbf{x}|\leq 1\}$, $A$ is a symmetric
nonnegative-definite $n\times n$ matrix, $\mathbf{b}\in \mathbb{R}^{n}$ and $%
\Pi (d\mathbf{x})$ is a measure on $\mathbb{R}^{n}$ such that
\[
\int_{\mathbb{R}^{n}}\min \{1,\langle \mathbf{x},\mathbf{x}\rangle \}\Pi (%
\mathbf{x})<\infty ,\,\,\Pi (\{\mathbf{0}\})=0.
\]%
Hence $\widehat{\mu }\left( \mathbf{y}\right) =e^{\psi (\mathbf{y})}$.
\end{em}

The density of $\Pi $ is known as the L\'{e}vy density and $A$ is the
covariance matrix. In particular, if $A=0$ (i.e. $A=(a_{j,k})_{1\leq j,k\leq
n}$, $a_{j,k}=0$) then the L\'{e}vy process is a pure non-Gaussian process
and if $\Pi =0$ the process is Gaussian. See \cite{gs1}-\cite{gs3}, \cite%
{McKean}, \cite{sato} for more information.

\end{document}